\documentclass[galaxies,article,submit,moreauthors,pdftex,10pt,a4paper]{mdpi}
\firstpage{1}
\articlenumber{x}
\doinum{10.3390/------}
\pubvolume{xx}
\pubyear{2018}
\copyrightyear{2018}
\history{Received: date; Accepted: date; Published: date}
\preto{\abstractkeywords}{\nolinenumbers}

\Title{The Astrochemistry Implications of Quantum Chemical Normal Modes Vibrational Analysis}

\Author{SeyedAbdolreza Sadjadi \orcidA{} \& Quentin Andrew Parker \orcidB{}}

\address[1]{
\quad Laboratory for Space Research and the Department of Physics, Faculty of Science, The University of Hong Kong, Pokfulam Road,  Hong Kong, China; ssadjadi@hku.hk, quentinp@hku.hk}

\abstract{Understanding the molecular vibrations underlying each of the unknown infrared emission (UIE) bands (such as those found at  3.3, 3.4, 3.5,  6.2, 6.9, 7.7, 
11.3, 15.8, 16.4, 18.9 $\mu$m) observed in or towards astronomical objects is a vital link to uncover the molecular identity of their carriers. This is usually done by 
customary classifications of normal mode frequencies such as stretching, deformation, rocking, wagging, skeletal mode, etc. A large literature on this subject exists 
and since 1952 ambiguities in classifications of normal modes via this empirical approach were pointed out by Morino and Kuchitsu \cite{MorinioJCP1952}. New 
ways of interpretation and analyzing vibrational spectra were sought within the theoretical framework of quantum chemistry 
\cite{KonkoliIJQC11998a,KonkoliIVJQC11998b}. Many of these methods cannot easily be applied \cite{KonkoliIVJQC11998b} to the large, complex molecular 
systems which are one of the key research interests of astrochemistry. In considering this demand, a simple and new method of analyzing and classifying the normal 
mode vibrational motions of molecular systems was introduced \cite{SadjadiApJ2015a}. This approach is a fully quantitative method of analysis of normal mode 
displacement vector matrices and classification of the characteristic frequencies (fundamentals) underlying the observed IR bands. Outcomes of applying such an 
approach show some overlap with customary empirical classifications, usually at short wavelengths. It provides a quantitative breakdown of a complex vibration (at longer wavelengths) into the contributed fragments like their aromatic or aliphatic components. In addition, in molecular systems \textbf{outside the classical models of chemical bonds and structures where the empirical approach cannot be applied, this quantitative method enables an interpretation of vibrational motion(s) underlying the IR bands.}  As a result, further modifications in \textbf {the structures (modeling) and the generation of the IR spectra (simulating) of} the UIE carriers, initiated by proposing a PAH model \cite{DuleyMonthyly1981PAH, AllamandolaApJS1989PAH}, can be implemented in \textbf{an efficient way. Here fresh results on the vibrational origin of the spectacular UIE bands based on astrochemistry molecular models, explored through the lens of the quantitative method applied to thousands of different vibrational motion matrices are discussed.} These results are important in the context of proto-planetary nebulae and planetary nebulae where various molecular species have been uncovered despite their harsh environments.}


\begin{document}

\setcounter{section}{0} 

\section{Introduction}
Detection in or towards various astronomical objects of broad emission infrared bands at certain wavelengths, i.e 3.3, 3.4, 3.5, 6.2, 6.9, 7.7, 8.7, 11.3, 15.8, 16.4 \& 
18.9 $\mu$m  led to the hypothesis that \textbf{organic molecules are the carriers} of these observed bands. \textbf {These are a diverse 
range of organic molecular compounds} from polyaromatic hydrocarbons(PAHs)\cite{DuleyMonthyly1981PAH,AllamandolaApJS1989PAH}, hydrogenated PAHs\cite{WagnerApJ2000HPAH}, 
nitrogenated PAHs \cite{HudginsASP2004NPAH}, nanodiamonds\cite{DuleyApJL2001Diamond}, oil fragments\cite{CataldoIJA2002}, quenched
carbonaceous composites (QCC)\cite{WadaQCC2006} and mixed aromatic-aliphatic organic nanoparticles(MAON)\cite{Kwoknature2011}. The main purpose of this work is to briefly discuss some new results of our on-going theoretical efforts and to explore the molecular vibrational origin of these mysterious UIE  bands \cite{SadjadiApJ2015a} through the lens of each of the proposed astrochemistry models above.

Most of the assignments of molecular vibrational motions and their responsible fragments or bonds are still based on an empirical approach
which can potentially  lead to incorrect interpretation of the molecular structure(s) of unknown carrier(s) of IR spectra \cite{MorinioJCP1952,KonkoliIJQC11998a,KonkoliIVJQC11998b}.  Our theoretical approach aims to provide complementary information to other methods of simulating fundamental, combination, overtone  \cite{CameronAnharmJCP2015, Hanson-HeineAnharmJPCA2012, MianiAnhamrJCP2000, GuntramVCIChemPhys2008} and  ro-vibrational\cite{DunhamRoVib1932, PavlyuchkoaRovib2015} bands of the Infrared spectra so generated.\

An example of an astrochemistry application of our methodology is our recent work on the vibrational origins of the 3.28 and 3.3 $\mu$m components of the 3.3 $
\mu$m feature known in the framework of the current PAH model \cite{SadjadiApJ2017}. The customary classification and assignment of the first component at 3.28$
\mu$m to the so-called \textit{bay} and the second at 3.3 $\mu$m, to the \textit{non-bay} -C-H stretching modes in PAH molecules \textbf{(for example there are two bay C-H bonds, each from one ring, in the phenanthrene molecule)} cannot adequately explain the observed 
wavelength difference and flux ratio of these two UIE features. This is due to the large and complex coupling between the stretching modes of aromatic -C-H  
bondings (bay and non-bay)\cite{SadjadiApJ2017}.

The primary empirical tool developed for interpreting and simulating the frequency and intensity of molecular fundamental infrared (IR) bands is the 
model of normal-mode vibration.  \cite{MorinioJCP1952} noted are no rules to exactly define the customary (empirical) classifications of normal mode 
frequencies as a result of bond stretching, deformation, rocking, wagging, skeleton mode, etc. The key role of having a well-defined system for assigning and 
classifying the vibrational motions in the interpretation of observed IR spectra was recently discussed by Tao et al\cite{TaoJCTC2018}. Their main argument is that 
normal modes are delocalized over all constituent atoms of a molecule. Hence, without any definite rules, assigning a vibration to a particular bond(s) or fragment(s) 
is a difficult and ambiguous task.

Although new mathematical methods to resolve this problem were introduced into quantum chemical models since 
1998\cite{KonkoliIJQC11998a,KonkoliIVJQC11998b}  these are not suitable methods to apply for analysis of vibrational motions of large molecules with complex 
structures\cite{KonkoliIVJQC11998b}, especially the types of molecules of particular interest in astrochemistry. The recent generalized subsystem vibrational 
analysis, developed by Tao et al\cite{TaoJCTC2018} might be a good framework for our purposes but the reliability of this model needs to be confirmed in different 
classes of molecules.   Clearly not all the possible 10$^{33}$ to 10$^{180}$ molecular species within the structural space of organic chemistry can be explored 
theoretically. Vibrational analysis and interpretation of IR spectra hence plays a key role in choosing the right bondings and fragments to modify in the astrochemistry 
molecular models and to inform a better understanding of the origin of the mysterious UIE bands. 

In this regard, we have developed and introduced a simple approach  for vibrational analysis based on  displacement vector analysis of normal modes 
\cite{SadjadiApJ2015a}. These are N $\times$ 3 matrices, where N is the number of atomic centers with a total number of 3N-6  normal modes for each molecule. 
We applied our methods to numerous examples of vibrational motions in simple and complex organic compounds composed of different types of chemical bonds, 
structures and sizes \cite{SadjadiApJ2015a,SadjadiApJ2015b,SadjadiJP2016,HsiaApJ2016,SadjadiApJ2017}. Our methodology provides consistent results 
compared to the more customary (empirical) approach and also in a fully quantitative fashion while additionally revealing important details of vibrational motions that 
are not recovered by the more traditional techniques. Another example is our detailed and comprehensive vibrational motion analysis of the complex correlation of a 
band's peak wavelength position of the out of plane bending mode vibration of aromatic -C-H bondings (OOPs) with the exposed edges (neighbour -C-H bondings) in 
the structure of PAH molecules\cite{SadjadiApJ2015b}. Such correlation was originally suggested as due to a monotonic increase in wavelength position by 
increasing the exposed edge \cite{HudginsApJL1999}.  

\section{Brief summary of our previous work}
Here we present {\bf a brief summary of our latest findings and understanding of the types of vibrational motions underlying UIE bands. This is based on our application of displacement vector analysis of vibrational normal modes to thousands of Raman and IR active modes of PAHs\cite{SadjadiApJ2017}} with different aliphatic and olefinic side chains\cite{SadjadiApJ2015a,SadjadiApJ2017} and MAON type molecules\cite{SadjadiJP2016}.
\textbf{In the customary empirical approach, the vibrational origin of an IR band is assigned to the vibration with strongest intensity. 
Unfortunately, the type of vibration and the responsible fragment(s)/bonds can not be correctly assigned in this way \cite{MorinioJCP1952,KonkoliIJQC11998a,KonkoliIVJQC11998b,TaoJCTC2018}. Neglecting the rest of the vibrational motions, including the Raman modes, will lead to incomplete or, in the case of long wavelength bands, an incorrect interpretation of the origin of an IR band.}  
Below is an itemized summary of our current findings. Thereafter our new results are presented for the first time.

\begin{itemize}[leftmargin=*,labelsep=5.8mm]
	        
\item In hydrocarbons within the realm of classical molecular structure only C-H stretching vibrations occur. This holds for any combination of different aromatic 
(sp$^{2}$), olefinic (sp$^{2}$) and aliphatic (sp$^{3}$) groups within a molecule.   Considering the low amplitude motions in our analysis we did not find the 
signatures of other types of vibrations in this spectral region.    

\item For the above types of molecules the aromatic (sp$^{2}$) C-H vibrations do not show any common wavelength coverage with aliphatic (sp$^{3}$)  stretching modes (non-overlap wavelength range). The vibrational motions of these two types of C-H bonds are reported to be uncoupled.    

\item Olefinic (sp$^{2}$) C-H stretching shows a common wavelength coverage with aromatic (sp$^{2}$) C-H vibrations (overlapped wavelength ranges ). However, vibrational motions of these two bond types occur independently without any couplings.

\item  Olefinic (sp$^{2}$) C-H stretching motion couples partly with aliphatic (sp$^{3}$) C-H vibrations. 

\item These vibrational characteristics of olefinic C-H stretching motions can glue the wavelength range of aliphatic (sp$^{3}$) and aromatic (sp$^{2}$) C-H stretching 
vibrations. In other words, they can make an indirect coupling between these two uncoupled and non-overlapped vibrational motions. This effect can be considered 
as one of the origins of the formation of a plateau in this spectral region.

\item The symmetric and asymmetric C-H stretching motions of methyl and methylene groups are highly coupled with common wavelength coverage of the resultant 
features. Thus vibrational motions of these two functional groups in this wavelength region are difficult to discriminate against under different UIE features.     

\item In neutral honeycomb PAH molecules, the stretching vibrations of C-H bonds in the bay and non-bay positions show considerable coupling in addition to their common wavelength coverage.    

\end{itemize}

\section{New results from this work}

The results presented here are from new quantum chemical calculations performed within density functional theory using mathematical tools to solve the Schrodinger equation for molecular systems.(\textbf{B3LYP and BHandHLYP functionals, in combination with a PC1 basis set family}). This model delivers an average error of 0.12-0.13 $\mu$m (within the wavelength range of 2 to 20 $\mu$m)  in reproducing the fundamental 
bands of laboratory gas phase IR spectra for the different classes of organic molecules \cite{SadjadiApJ2015a}. The effects of anharmonicity in vibrations and 
rotational-vibrational couplings on final IR bands wavelength positions are considered by applying the Laury et al \cite{LauryJCC2012} scheme of double-scale 
factors for harmonic normal mode calculations.    

\subsection{The 3 $\mu$m region}
\subsubsection{Models: PAH, PAHs with side groups, Aliphatic hydrocarbons and MAONs}	
Although this part of the emission spectra seems to be well explored in terms of the types of vibrational motions in the framework of cited astrochemistry models, 
plateau formation, the flux ratio of bands used to estimate the aliphatic/aromatic contents of UIE carriers, the long wavelength UIE features (such as the 3.51$\mu$m 
peak assigned to nano-diamond structures) and the relationship of these features to the other UIEs at different IR wavelengths  have remained unsolved issues. A 
comprehensive vibrational analysis and classification of normal modes will play a key role in addressing these issues. Specifically, we found that by keeping the molecular structure of organic compounds within the framework of classical Lewis model of bonding and structure the classifications 
and assignments of normal modes are not simple and definite as implied from the customary (empirical) approach. Perhaps classifications, assignments, and 
interpretation of vibrational motions underlying IR bands within the empirical approach would be ambiguous for non-classical molecular structures such as an
amorphous type complex hydrocarbon molecule shown in Figure~\ref{fig1}. 
Here a modern and advanced molecular chemical bonding and structure theory such as the 
quantum theory of atoms in molecules should be recalled to assist the interpretation \cite{Baderbook1990,Gillespiebook2001,MattaBoydBook2007}.\\


\subsection{The 6-10 $\mu$m region}
\subsubsection{Model used: PAH}

Figure~\ref{fig2} summarizes the results of our new, quantitative vibrational analysis followed by statistical manipulations within the wavelength step size of 0.1$
\mu$m, performed on all normal modes of a set of 70 neutral honeycomb PAH molecules with the size of 6 to 136 carbon atoms 
\cite{SadjadiApJ2015b,SadjadiApJ2017}. Based on this new mathematical analysis we find that:

\begin{itemize}[leftmargin=*,labelsep=5.8mm]
	
\item All normal mode vibrations (IR and Raman active modes) within this wavelength range occur in the symmetry plane of the PAH molecule. They are all in-plane 
modes. We did not detect any low amplitude out of plane modes through our quantitative analysis.    

\item All vibrations are reported as highly coupled modes. This includes the 6.2 $\mu$m feature described as an uncoupled C-C stretching band (zone (a) in 
Figure~\ref{fig2}). Only the wavelength range of 8.6 to 8.9 $\mu$m is found to contain pure aromatic C-H in-plane bending modes (zone (c) in Figure~\ref{fig2}).

\textbf{These pictures of vibrations, derive and assigned purely from theory} are in very good agreement with the results of experimental spectroscopy reported by Jobelin et al \cite{JoblinAA1994} from IR spectra of numbers of PAH molecules. \textit{This agreement with experimental work demonstrates the reliability of our quantitative vibrational analysis methodology which is a key 
point to exploring the origin of UIE bands at longer wavelengths.} 

\item In Figure \ref{fig2} we have labeled and described these vibrations as coupled C-C stretching and C-H in-plane modes. These are commonly used labels. 
These types of motions also occur at longer wavelengths in PAH molecules. We explain later a possible way of discriminating such vibrations and assigning 
them to intrinsic fragments. It should be mentioned that the comprehensive vibrational analysis of this 6-10 $\mu$m region and the 3 $\mu$m part of the spectra has 
led us to suggest that the olefininc carrier of the 6 $\mu$m UIE feature observed in the famous proto-planetary red rectangle nebulae HD44179 \cite{HsiaApJ2016}. 
This feature was previously assigned to CO molecule.    
 
\end{itemize}

\subsection{The 11-15 $\mu$m region}
\subsubsection{Mode usesl: PAH}

For neutral planar honeycomb PAH molecules, this wavelength range contains two major types of motions. They are either in-plane or out of molecular plane 
vibrations. We did not detect any vibrational motions composed of both types of these vibrations. Each of these two major types of vibrations has their own 
subdivision vibrations, where the out of plane bending mode of aromatic C-H bondings is one of the well-known categories. The origin of each UIE feature in this 
wavelength range is complex and cannot be assigned by one label.  By filtering all coupled vibrational motions, It is found that the pure uncoupled C-H out of plane bending modes vibrate at different frequencies, independent of their classical exposed edge origin (\cite{SadjadiApJ2015b}).  

\subsection{Skeletal modes}
\subsubsection{Model used: MAONs}

One of the ambiguous concepts in vibrational analysis is the concept of the skeletal mode (or skeleton mode). Usually, this refers 
to the vibrational motion of a part of the molecules composed of elements heavier than hydrogen, such as the C-C stretching mode in PAH 
molecules\cite{AllamandolaSkeletal1985}. Such a picture is also satisfied 
by the other types of vibrational motions over both short and long wavelength ranges in 
different classes of organic molecules.  This concept is becoming more confusing by introducing heteroatoms like nitrogen, oxygen, sulphur, etc... into the structure 
of organic molecules.

Here we provide an alternative way to define skeletal mode by considering all atoms not only heavy ones. We chose the MAON model\cite{Kwoknature2011}, because of its 3D structural diversity. The results of vibrational analysis of 17,257 normal modes (IR and Raman active modes) of 56 MAON type molecules are plotted in Figure \ref{fig3} from 2 to 30 $\mu$m. All MAONs were composed of benzene, methyl, and methylene classical functional groups. The statistical analysis has been carried out within the wavelength interval of 0.1 $\mu$m. The simulated IR signature of these MAONs can be found in \cite{SadjadiJP2016}. Such a statistical plot (Figure~\ref{fig3}) can be obtained for any class of organic molecule by applying our methodology \cite{SadjadiApJ2015a}. 

\begin{itemize}[leftmargin=*,labelsep=5.8mm]
	
\item In Figure~\ref{fig3}, the smallest number of total atoms that contributes to the vibrations obtained within a 0.1 $\mu$m interval is plotted against wavelength. 
On average this number shows a sudden increase from 14.4 $\mu$m towards longer wavelengths. This enables us to split the IR spectra of MAONs into two major 
regions (Figure\ref{fig3}). We suggest that the long wavelength section is labeled as the skeletal mode section. Here the vibrational motions are delocalized in nature. 
The short wavelengths section which contains localized vibrations (i.e. a smaller number of atoms are involved) is labeled as bonding modes.     

\end{itemize}
  
\begin{figure}[H]
	\centering
	\includegraphics[width=10 cm]{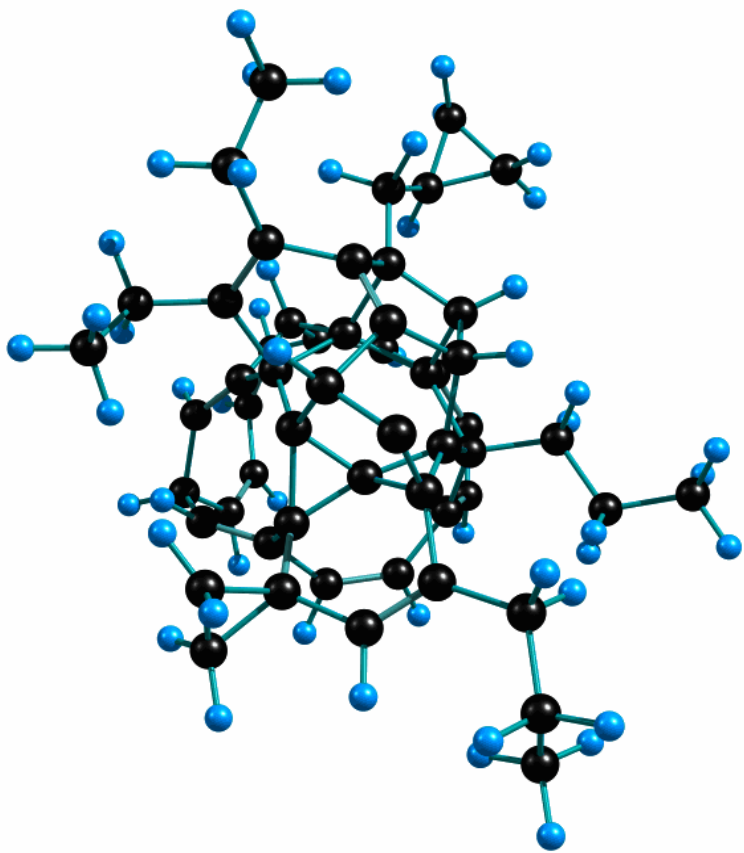}
	\caption{Local minimum geometry of an amorphous type complex hydrocarbon molecule (C$_{55}$H$_{52}$) at B3LYP/PC1 model. The molecule is composed of non-classical core and classical side groups.}
	\label{fig1}
\end{figure}

\begin{figure}[H]
	\centering
	\includegraphics[width=10 cm]{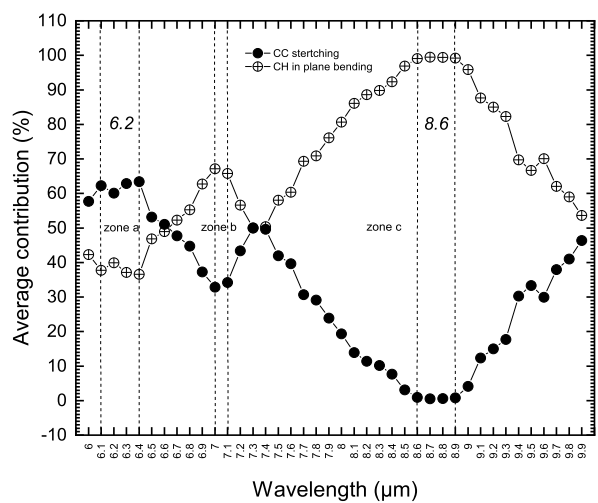}
	\caption{The newly calculated changes in average contributions of aromatic C-C stretching and C-H in plane bending modes in neutral honeycomb PAH molecules.}
	\label{fig2}
\end{figure}

\begin{figure}[H]
\centering
\includegraphics[width=10 cm]{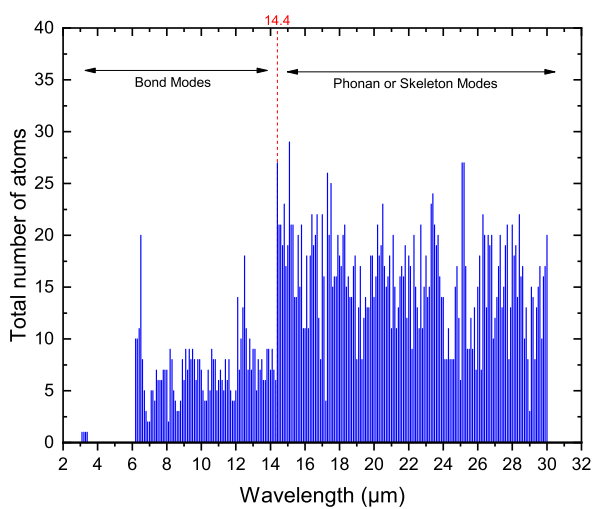}
\caption{A new plot presented for the first time of statistical analysis done on the results of displacement vector matrix analysis of 17,257 normal modes (all Raman and IR active modes) in 56 different mixed aromatic-aliphatic organic nanoparticles (MAON) type molecules. Statistical analysis performed within the wavelength interval of 0.1 $\mu$m from 2 to 30 $\mu$m. The abrupt change in the total number of participant atoms occurs at a wavelength of $\sim$14.4 $\mu$m. Two regimes of vibrational motions are distinguishable as bond(bonding) and skeletal(skeleton) modes in MAON molecules.}
	\label{fig3}
\end{figure}

\section{Conclusions}

Despite the harsh conditions, the rich chemistry of proto- and planetary nebulae is seen in their strong UIE bands.  \textbf{Here, new quantitative and comprehensive results of displacement vector analysis of normal modes are reported for PAH molecules in the range of (6-10 $\mu$m)  - see Figure \ref{fig2}). The C-C stretching, C-H bending coupling diagram (Figure 2), provides theoretical explanations for laboratory experimental observations as well as describing UIE band origin in a quantitative way in light of the PAH model. The analysis of $>$17,000 vibrational normal modes of MAONs with different numbers of atoms and structures within the range of (2-30 $\mu$m) are presented for the first time (Figure \ref{fig3}). This shows a boundary at 14.4 $\mu$m between two regimes of complex molecular motions and provides clues to the bond(s) and fragment(s) responsible for different IR features.  With such valuable quantitative information, modeling the structure of unknown UIE carries and further computations on their spectroscopic properties are performed effectively. With the correct description of the molecular vibrations underlying the UIE bands, we can get closer to explaining the mechanism of the formation of another spectacular feature of UIE bands, i.e. the plateaus at 8, 12, 17 and 21 $\mu$m. It should be emphasized that this type of molecular vibrational motion analysis is not restricted, like customary empirical approaches, to the molecules with classical bond types, fragments, and functional groups. The complex molecular vibrations in Figure \ref{fig1} with non-classical bond types and structure (and unlimited numbers of such molecules), can be analyzed. This will push the modeling of UIE carriers to include new species with exotic properties.}

Hence, computational quantum chemistry provides vital theoretical resources to discover the molecular carriers of these mysterious bands. This is currently performed by modeling the spectroscopic properties of large numbers of complex organic molecules. The calculated data is then processed via vibrational analysis for interpretation of the origin of UIE features. This provides valuable information linking the spectra to molecular structure. This assists further 
modifications of the proposed astrochemistry molecular models. Although the link between molecular structure and IR features is a complex relationship 
and sometimes confusing, our present results of vibrational analysis on thousands of normal modes on different classes of organic molecules can 
contribute significantly in addressing the vibrational origin of UIE bands and their unknown molecular carriers.    


\vspace{6pt}



\acknowledgments{We thank Prof.Sun Kwok for his previous leadership of the LSR Astrochemistry program. We are also grateful to colleagues Dr. Yong Zhang and Dr. Chih Hao Hsia for useful discussions. The computations were performed using the computing facilities provided by the University of Hong Kong.}

\authorcontributions{\textit{\textbf{The first author undertook all the mathematical modelling based on the existing collaboration and wrote the bulk of the manuscript. The second author as supervisor provided some oversight, critique, discussion, vetting and re-writing of certain aspects of the paper.}}}
\conflictsofinterest{The authors declare no conflict of interest.} 
\reftitle{References}




\end{document}